# Cobalt-56 γ-ray emission lines from the type Ia supernova 2014J

E. Churazov[1,2], R. Sunyaev[1,2], J. Isern[3], J. Knödlseder[4,5], P. Jean[4,5], F. Lebrun[6], N. Chugai[7], S. Grebenev[1], E. Bravo[8], S. Sazonov[1,9] & M. Renaud[10]

[1]Space Research Institute (IKI), Profsouznaya 84/32, Moscow 117997, Russia; [2]Max Planck Institute for Astrophysics, Karl-Schwarzschild-Strasse 1, 85741 Garching, Germany; [3]Institute for Space Sciences (ICE-CSIC/IEEC), 08193 Bellaterra, Spain; [4]Universite´ de Toulouse, UPS-OMP, IRAP, Toulouse, France; [5]CNRS, IRAP, 9 Avenue colonel Roche, BP 44346, F-31028 Toulouse Cedex 4, France; [6]APC, Universite´ Paris Diderot, CNRS/IN2P3, CEA/Irfu, Observatoire de Paris, Sorbonne Paris Cite´, 75205 Paris Cedex 13, France; [7]Institute of Astronomy of the Russian Academy of Sciences, 48 Pyatnitskaya Street, 119017 Moscow, Russia; [8]ETSAV, Universitat Politecnica de Catalunya, Carrer Pere Serra 1-15, 08173 Sant Cugat del Valles, Spain; [9]Moscow Institute of Physics and Technology, Institutsky pereulok 9, 141700 Dolgoprudny, Russia; [10]LUPM, Universite´ Montpellier 2, CNRS/IN2P3, CC 72, Place Euge` ne Bataillon, F-34095 Montpellier Cedex 5, France.

**A type Ia supernova is thought to be a thermonuclear explosion of either a single carbon–oxygen white dwarf or of a pair of merging white dwarfs. The explosion fuses a large amount of radioactive $^{56}$Ni (refs 1–3). After the explosion, the decay chain from $^{56}$Ni to $^{56}$Co to $^{56}$Fe generates γ-ray photons, which are reprocessed in the expanding ejecta and give rise to powerful optical emission. Here we report the detection of $^{56}$Co lines at energies of 847 and 1,238 keV and a γ-ray continuum in the 200–400 keV band from the type Ia supernova 2014J in the nearby galaxy M82. The line fluxes suggest that about $0.6 \pm 0.1$ solar masses of radioactive $^{56}$Ni were synthesized during the explosion. The line broadening gives a characteristic mass-weighted ejecta expansion velocity of $10,000 \pm 3000$ km/s. The observed γ-ray properties are in broad agreement with the canonical model of an explosion of a white dwarf just massive enough to be unstable to gravitational collapse, but do not immediately exclude more complicated merger scenarios, which fuse comparable amount of $^{56}$Ni.**

The detailed physics of the explosion of type Ia supernovae (for example deflagration or detonation) and the evolution[4,5] of a compact object towards explosion remain a matter of debate[6–9]. In a majority of models, the ejecta are opaque to γ-ray lines during first 10–20 days after the explosion (because of Compton scattering). At later times, the ejecta become progressively more transparent and a large fraction of γ-rays escapes. This leads to a robust prediction[10] of γ-ray emission from type Ia supernovae after few tens of days, dominated by the γ-ray lines of $^{56}$Co. Such emission has been observed before: the down-scattered hard X-ray continuum from supernova (SN) 1987A in the Large Magellanic Cloud was seen half a year after the explosion[11,12], and γ-ray lines of $^{56}$Co were detected several months later[13,14]. That was a core-collapse (type II) supernova, in which the cause of the explosion is completely different from that of type Ia supernovae. Type Ia events, despite being intrinsically brighter, are more rare than core-collapse supernovae, and before

SN 2014J there was not one close enough to detect. The recent type Ia SN 2011fe at a distance of $D \approx 6.4$ Mpc yielded only an upper limit on the $^{56}$Co line flux[15].

SN 2014J in M82 was discovered[16] on January 21, 2014. The reconstructed[17] date of the explosion is January 14.75 UT. This is the nearest type Ia supernova to be detected in at least four decades, at the M82 distance of $D \approx 3.5$ Mpc (ref. 18). The European Space Agency satellite INTEGRAL[19] started observing SN 2014J in 2014, from January 31 to April 24 and from May 27 to June 26. We use the INTEGRAL data taken between days 50 and 100 after the explosion, the period when the expected flux from γ-ray lines of $^{56}$Co is close to the maximum[10]. This set of observations by the SPI and ISGRI/IBIS instruments on board INTEGRAL has been analysed, excluding periods of strong solar flares, which cause large variations in the instrumental background (Methods and Extended Data Fig. 1). The spectrum derived assuming a point source at the position of SN 2014J is shown in Fig. 1 using red and blue points for SPI and ISGRI, respectively.

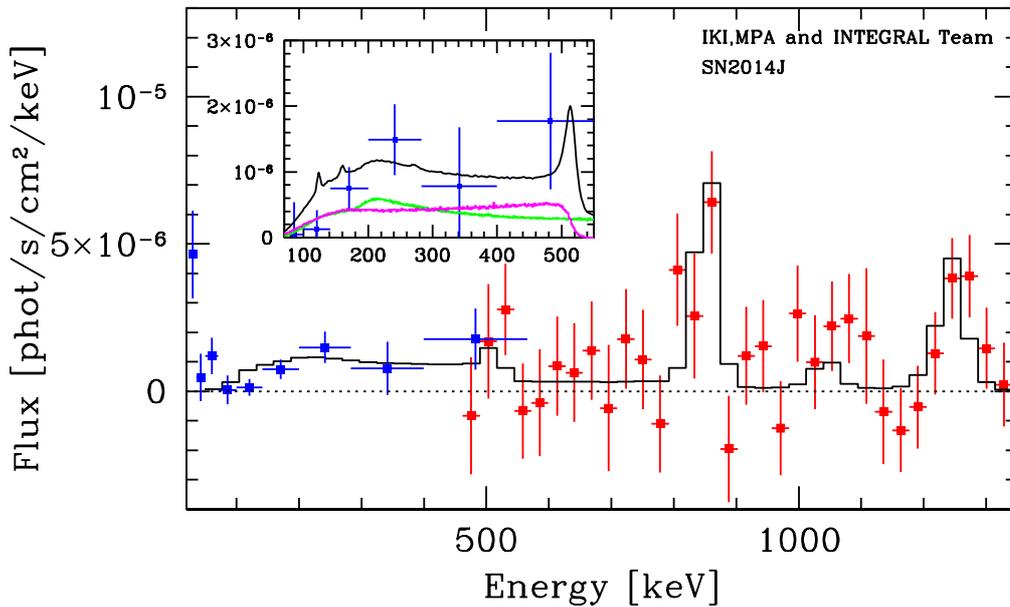

***Figure 1 Gamma-ray lines from Co decay at 847 and 1,238 keV in the spectrum of SN 2014J.*** *The spectrum was obtained by INTEGRAL between days 50 and 100 after the outburst. Red and blue points show SPI and ISGRI/IBIS data, respectively. The flux below 60 keV is dominated by the emission of M82. The black curve shows a fiducial model of the supernova spectrum for day 75 after the explosion. Inset, lower-energy part of the spectrum (black). The expected contributions of three-photon positronium annihilation (magenta) and Compton down-scattered emission from 847 and 1,238 keV lines (green) are also shown. All error bars, 1 s.d.*

The model spectrum is binned similarly to the observed supernova spectrum. The signatures of the 847 and 1,238 keV lines are clearly seen in the spectrum (along with tracers of weaker lines of $^{56}$Co at 511 and 1,038 keV). The low-energy (<400 keV) part of the SPI spectrum is not shown because of possible contamination due to off-diagonal response of the instrument to higher-energy lines. At these energies, we use ISGRI/IBIS data instead (Methods).

By varying the assumed position of the source and repeating the flux-fitting procedure using SPI data (Methods) we construct a 40° × 40° image of the signal-to-noise ratio in the 800–880 and 1,200–1,300 keV energy bands (Fig. 2). SN 2014J is detected at 3.9 s.d. and 4.3 s.d. in these two bands, respectively. These are the highest peaks in both images.

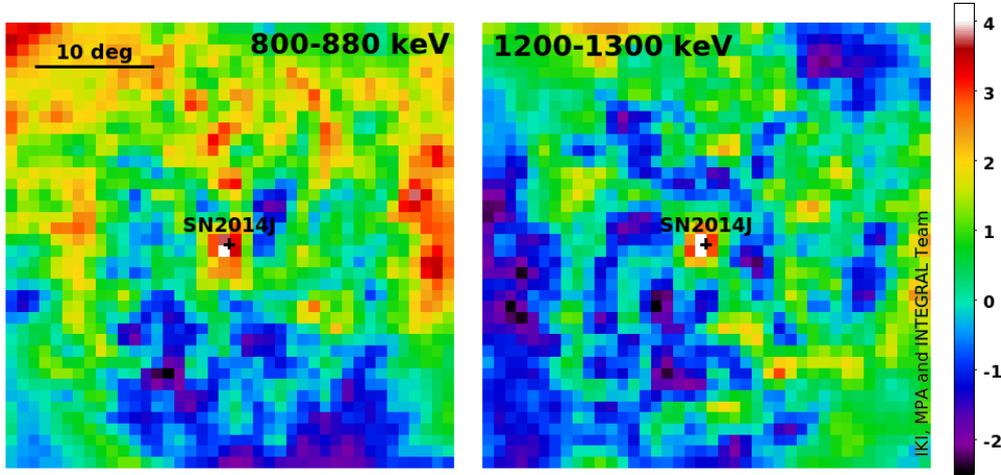

*Figure 2 Signatures of $^{56}$Co lines at 847 and 1,238 keV in SPI images. The broad energy bands 800–880 keV and 1,200–1,300 keV are expected to contain the flux from $^{56}$Co lines, accounting for the expected broadening and shifting due to the ejecta expansion and opacity effects. The source is detected at 3.9 s.d. and 4.3 s.d., respectively, in these two bands. Colours show the signal-to-noise ratio at a given position. Data courtesy of IKI, MPA and the INTEGRAL team.*

The images obtained by ISGRI at lower energies (100–600 keV) during the observations of SN 2014J and in October–December 2013, that is, a few months before the SN 2014J explosion (see Methods for the details of the earlier observation), are shown in Fig. 3. An inspection of images in the 25–50 keV band shows that the fluxes observed in 2013 and 2014 are similar, whereas at higher energies (>100 keV) there is excess at the position of SN 2014J only in the 2014 data. Previous ISGRI observations of this field in 2009–2012, with a total exposure of about $6 \times 10^6$ s, revealed no significant signal at energies >50 keV from M82[20].

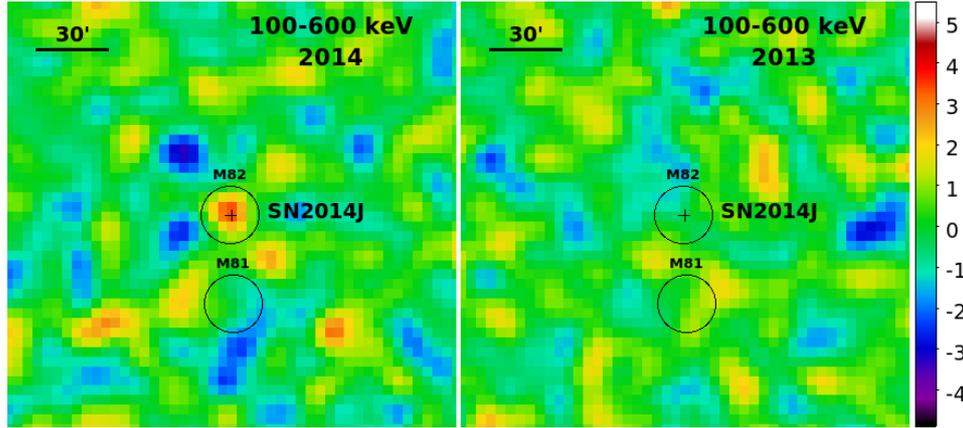

*Figure 3 Appearance of a new hard (100–600 keV) X-ray source at the position of SN 2014J. In the ISGRI image of the M82 field taken in 2013 the source is absent. Colours show the signal-to-noise ratio at a given position. SN 2014J is detected in this image at ~3.7 s.d.*

A combination of imaging and spectral analysis provides robust evidence of γ-ray emission from SN 2014J. As expected, much of the signal comes from the 800–900 and 1,200–1,300 keV bands, where two prominent lines of $^{56}$Co should be. The best-fitting parameters (flux, energy and broadening) of those two lines are given in Extended Data Table 1. The fluxes at 847 and 1,238 keV, respectively $(2.34 \pm 0.74) \times 10^{-4}$ and $(2.78 \pm 0.74) \times 10^{-4}$ photons s$^{-1}$ cm$^{-2}$, were corrected for the fraction of γ-ray photons escaping the ejecta (this fraction is ~60%; Methods), yielding an estimate of $M_{Co} = (0.34 \pm 0.07)M_\odot$ for the total $^{56}$Co mass at day 75 ($M_\odot$, solar mass). Correcting this value for the secular evolution of the $^{56}$Co mass in the decay chain $^{56}$Ni → $^{56}$Co → $^{56}$Fe, provides an estimate of the initial $^{56}$Ni mass: $M_{Ni} = (0.61 \pm 0.13)M_\odot$.

An independent estimate of the $^{56}$Ni mass can be obtained from the bolometric light curve. The simplest approach is based on the assumption[21] that the bolometric luminosity at the maximum is approximately equal to the power of the radioactive decay at this moment. For SN 2014J, the maximum bolometric luminosity is ~$1.1 \times 10^{43}$ erg s$^{-1}$, attained on day 17.7 after the explosion[22], assuming interstellar extinction in V filter band $A_V = 1.7$. A thorough analysis of extinction[23] yields $A_V = 1.85 \pm 0.11$ mag. This implies a $^{56}$Ni mass of $(0.42 \pm 0.05)\ M_\odot$. This is marginally consistent with the estimates based on the γ-ray emission lines, which is not surprising, given the qualitative nature of this estimate and the large and complicated extinction in the direction of SN 2014J. A more direct test is the comparison of γ-ray and bolometric optical, ultraviolet and infrared luminosities at day 75. The latter is estimated to be ~1.1 10$^{42}$ erg s$^{-1}$ (Methods). The total energy released during decay of the $^{56}$Co isotope[24] is split between neutrinos (~0.8 MeV), kinetic energy of positrons (~0.12 MeV) and γ-rays (~3.6 MeV). In our fiducial model a fraction $f \approx 0.77$ of the luminosity in γ-rays escapes the ejecta at day 75. The remaining fraction, $1 - f \approx 0.23$, is deposited in the ejecta (ignoring bremsstrahlung radiation losses by electrons). Adding the kinetic energy of positrons and 23% of the γ-ray luminosity produced by $0.34 M_\odot$ of $^{56}$Co yields an estimate of $\sim 1.1 \times 10^{42}$ erg s$^{-1}$ for the rate of energy deposition in the ejecta in good agreement with the optical data. The same model predicts that $\sim 3.3 \times 10^{42}$ erg s$^{-1}$ escape the ejecta in the form of hard X-rays and γ-rays. The observed luminosities of the 847 and 1,238 keV lines are $\sim 4.7 \times 10^{41}$ and $\sim 8.1 \times 10^{41}$ erg s$^{-1}$, respectively.

The emergent lines are expected to be broadened and blueshifted because of ejecta expansion and the opacity effects (Methods and Extended Data Fig. 3). Both effects are indeed observed (Fig. 4). The mean blueshift, averaged over both lines, corresponds to a velocity of $V_{\text{Shift}} = -3{,}100 \pm 1{,}100$ km s$^{-1}$, and the broadening (root mean squared line-of-sight velocity) is $\sigma_\gamma = 4{,}100 \pm 960$ km s$^{-1}$. These values are broadly consistent with expectations of our fiducial model (Methods).

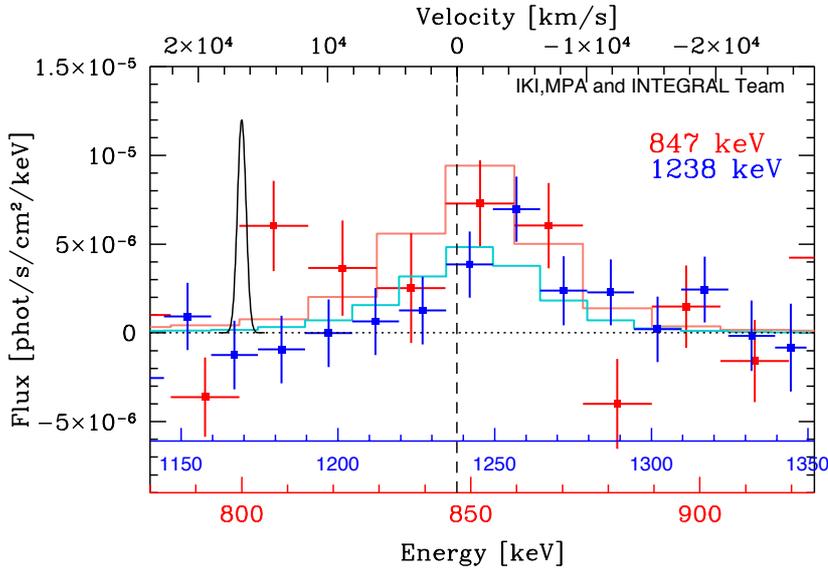

*Figure 4 Broadening of the 847 and 1,238 keV lines.* Red points show the SPI spectrum in the 720–920 keV range. The red histogram shows the line profile in the fiducial model. The blue points show the SPI spectrum of the 1,238 keV line. For comparison, a Gaussian line at 800 keV with a width corresponding to the SPI intrinsic energy resolution is shown with a black line. Both observed lines are clearly broadened. The upper axis shows the velocity needed to shift the line to a given energy. All error bars, 1 s.d.

Finally, at lower energies (100–400 keV) the emerging flux is dominated by Compton scattering of the 847 and 1,238 keV photons and the ortho-positronium continuum from positron annihilation (Fig. 1 inset and Extended Data Fig. 4). The predicted 100–400 keV flux is consistent with INTEGRAL data. Roughly half of the signal comes from the ortho-positronium continuum, suggesting that at day 75 the positrons produced in 19% of the $^{56}$Co decay are thermalized in the ejecta and annihilate via positronium formation. Below 100 keV, the emission declines strongly owing to photoabsorption.

It should be possible to derive stronger constraints from the combined analysis of the whole spectrum. As a first step, we constructed a three-parameter ($M_{Ni}$; ejecta mass, $M_{Ej}$; characteristic velocity, $V_e$) model that is capable of reproducing the main observables in the spectrum: the γ-ray line flux, the line broadening and the continuum flux below 511 keV (Methods). The model assumes spherical symmetry, complete mixing of all elements over the entire ejecta, and an exponential density profile[25,26]: $\rho \propto e^{-v/V_e}$. All three parameters are treated as independent. The level of mixing can in principle be determined from the time evolution of the γ-ray flux. For example, early appearance of hard X-ray emission from core-collapsed SN 1987A clearly demonstrated that Co is mixed[27] over the ejecta. A Monte Carlo code follows the propagation of the γ-ray photons through the ejecta and accounts for scattering and photoabsorption of photons and annihilation of positrons (Methods).

We vary $M_{Ni}$, $M_{Ej}$ and $V_e$ over a wide range, calculate the expected emergent spectrum and compare it (in terms of $\chi^2$ test) with the observed spectrum. This procedure yields the following best-fit parameters and 1 s.d. confidence intervals for the individual parameters: $M_{Ni} = 0.56^{+0.14}_{-0.06} M_\odot$, $M_{Ej} = 1.2^{+1.9}_{-0.5} M_\odot$, $V_e = 3000 \pm 800$ km s$^{-1}$. In this model the mass-weighted root-mean-squared velocity of the ejecta is $\approx \sqrt{12} V_e = 10,000 \pm 3,000$ km s$^{-1}$. (Methods and Extended Data Fig. 5). The confidence contours in the $M_{Ni}$–$M_{Ej}$ plane are elongated such that the highest and lowest allowed values of $M_{Ni}$ respectively correspond to the highest and lowest values of $M_{Ej}$.

In more realistic models, based on calculations of explosive nucleosynthesis, the parameters are not independent and the distribution of elements over the ejecta can vary strongly. We therefore compared the expected spectra for several representative models of type Ia supernovae at day 75, scaled to the distance of M82, with the data. The list of models and corresponding $\Delta\chi^2$ values with respect to the null hypothesis of no source are given in Extended Data Table 2. Remarkably, the canonical model of type Ia supernovae, W7[2], provides the best description of the SN 2014J spectrum, with $\Delta\chi^2 = 54.4$. The pure-detonation model DETO[28] produces too much $^{56}$Ni and can be reliably rejected. The sub-Chandrasekhar model HeD6[29] instead produces a γ-ray flux that is too low and therefore can also be rejected. Our best-fitting three-parameter model (3PAR), and delayed-detonation models [DD4[30] and DDT1p (E.B. *et al.*, manuscript in preparation)], designed to approximately match SN 2014J properties in the visual band, offers a gain in $\Delta\chi^2$ nearly as good as W7. From the standpoint of purely statistical errors, W7 performs significantly better than other models. However, given the inevitable assumptions and simplification in each model, and allowing for possible systematic uncertainties, this group of models cannot be rejected.

Overall, the good agreement with the canonical models shows that in γ-rays SN 2014J looks like a prototypical type Ia supernova, even though strong and complicated extinction in the optical band makes the overall analysis challenging.

**Acknowledgements** This work was based on observations with INTEGRAL, an ESA project with instruments and a science data centre funded by ESA member states (especially the principal investigator countries: Denmark, France, Germany, Italy, Switzerland and Spain) and with the participation of Russia and the United States. We are grateful to the ESA INTEGRAL team and E. Kuulkers for their prompt reaction to the SN 2014J event. E.C., R.S. and S.G. wish to thank the Russian INTEGRAL advisory committee for allocating an additional $10^6$ s of time from a regular program to SN 2014J observations. E.C., R.S., S.G., S.S. are partly supported by grant No. 14−22−00271 from the Russian Scientific Foundation. J.I. is supported by MINECO-FEDER and Generalitat de Catalunya grants. The SPI project has been completed under the responsibility and leadership of CNES, France. ISGRI has been realized by CEA with the support of CNES. The authors thank Peter Höflich, Ken Nomoto, and Stan Woosley for making available their supernova explosion models HED6, W7, and DD4.

## METHODS

**Observations**

INTEGRAL is an ESA scientific mission dedicated to fine spectroscopy and imaging of celestial γ-ray sources in the energy range from 15 keV to 10 MeV. The INTEGRAL data used here were accumulated during revolutions 1391–1407 (ref. 31), corresponding to the period ~50–100 days after the explosion (proposals: 1170002, PI: R.S.; 1140011, PI: J.I.; 1170001, public). Observations were performed in a standard 5 × 5 pattern around the nominal target location: one source on-axis pointing, 24 'off-source pointings', with a 2.17° step. During off-source pointings, the source remains well within the INTEGRAL field of view. Periods of very high and variable background due to solar flares were omitted from the analysis (Extended Data Fig. 1). The total exposure of the clean data set is ~2.6 Ms.

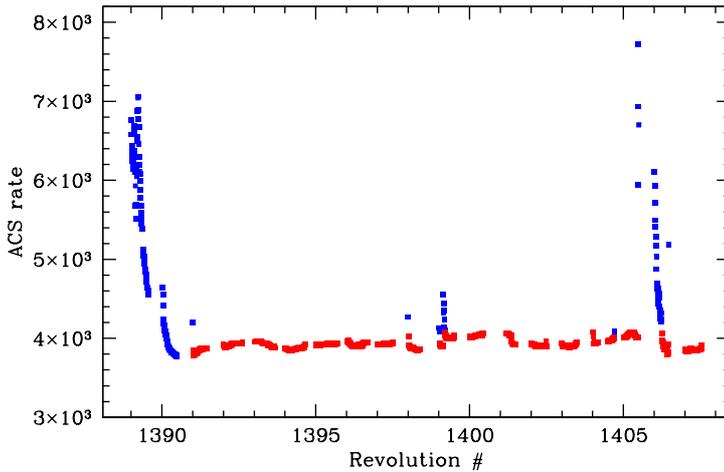

*Extended Data Figure 1*

*Variations in the particle background during INTEGRAL observations. Anti-coincidence-system count rate is shown as a function of time, expressed through the revolution number. Periods of very high and variable background (shown in blue) due to solar flares were omitted from the analysis. Periods of quiescent background (red) were used to derive the spectrum of SN 2014J.*

**SPI data analysis**

SPI[32] is a coded-mask germanium spectrometer on board INTEGRAL. The instrument consists of 19 individual Ge detectors, and has a field of view of ~30° (at zero response), an effective area of ~70 cm$^2$ at 0.5 MeV and an energy resolution of ~2 keV. The effective angular resolution of SPI is ~2°. During SN 2014J observations, 15 out of 19 detectors were operating, resulting in slightly reduced sensitivity and imaging capabilities compared with the initial configuration. The data analysis follows the scheme implemented for the analysis of the Galactic Centre positron annihilation emission[33,34]. For each detector, a linear relation between the energy and the channel number was assumed and calibrated (separately for each orbit), using the observed energies of background lines (Extended Data Fig. 2) at ~198, 438, 584, 882, 1,764, 1,779, 2,223 and 2,754 keV. These bright lines are identified[35] with known nuclear transitions in Ge and other elements, activated by heavy particles, bombarding the SPI detector. For our analysis, we used a combination of single and pulse-shape-discriminator (PSD) events[32], and treated them in the same way.

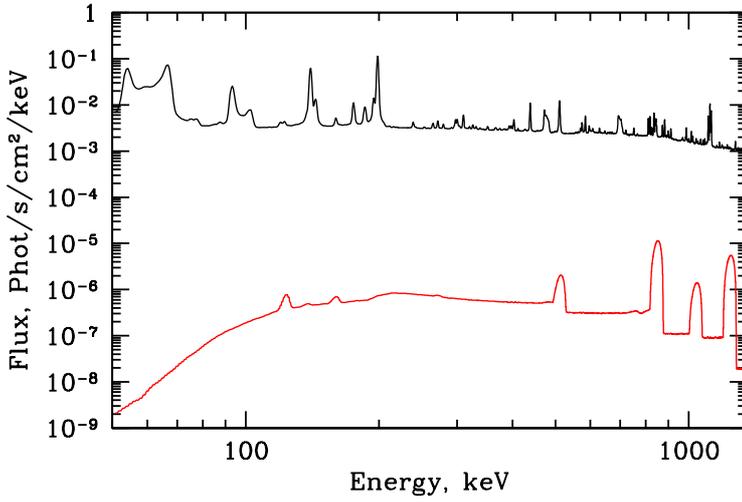

***Extended Data Figure 2***

*Comparison of the SPI background spectrum and the expected type Ia supernova emission.* Typical quiescent background (black) and supernova model (red, convolved with SPI energy resolution) spectra.

The count rate from the supernova at energy $E$, $S(E)$, and the background rates in individual detectors, $B_i(E, t)$, were derived from a simple model of the observed rates, $D_i(E, t)$, in individual SPI detectors, where $i$ is the detector number and $t$ is the time of observation with a typical exposure of 2,000 s: $D_i(E, t) \approx S(E) \times R_i(E, t) + B_i(E, t) + C_i(E)$. Here $R_i(E, t)$ is the effective area for the $i$th detector, as seen from the source position in a given observation, and $C_i(E)$ does not depend on time. The background rate is assumed to be linearly proportional to the Ge detectors' saturated event rate, $G_{Sat}(t)$, above 8 MeV, averaged over all detectors; that is, $B_i(E, t) = \beta_i(E) G_{Sat}(t)$. The coefficients $S(E)$, $\beta_i(E)$ and $C_i(E)$ are free parameters of the model and are obtained by minimizing $\chi^2$ for the entire data set.

Even though the number of counts in individual exposures is low, it is still possible to use a plain $\chi^2$ approach as long as the errors are estimated using the mean count rate and the total number of counts in the entire data set is large[36]. The linear nature of the model allows for straightforward estimation of statistical errors.

As an example, we consider SPI data in the 1,200–1,300 keV range, containing the 1,238 keV line of $^{56}$Co. For the entire data set there are in total 14,340 flux measurements by individual SPI detectors with a typical exposure time of ~2,000 s. Fitting this data set with a constant background model (that is, $D_i(E, t) \approx C_i(E)$) yields the reduced $\chi^2_r = 1.31$ per degree of freedom. Adding a $B_i(E, t) = \beta_i(E) G_{Sat}(t)$ term to the model reduces $\chi^2_r$ to 0.998, consistent with expectations for photon counting noise. Further adding one more parameter—the flux of a source at the position of SN 2014J—lowers $\chi^2$ by $\Delta\chi^2 = 16.4$, corresponding to a ~4 s.d. detection. For comparison, for the 600–800 keV band, which does not contain strong lines, the improvement in $\chi^2$ by adding the source flux to the model is only 0.60, giving no significant evidence for SN 2014J emission.

Despite its proximity, SN 2014J is still an extremely faint source in γ-rays. Although precise measurements of line fluxes are challenging, a combination of spectral and imaging information makes our results very robust. To further test the possible influence of variable background on our results, we repeated the calculation of the SN 2014J spectrum while dropping the $B_i(E, t)$ term from the model. The line fluxes changed by ~3% compared with our more elaborate baseline background model. However, the wings of the lines, where the background is ~$10^4$ times higher than the source, can still suffer from the residual background features. The uncertainty in the distance $D = 3.53 \pm 0.26$ Mpc (ref. 18) causes an additional ~15% uncertainty in the flux and, therefore, in the estimated mass of radioactive Ni and Co.

**ISGRI/IBIS data analysis**

The primary imaging instrument on board INTEGRAL is IBIS[37], which is a coded-mask aperture telescope with the CdTe-based detector ISGRI[38]. It has higher sensitivity to continuum emission than SPI in the 20–300 keV range and has a spatial resolution of ~12′. We note here that neither ISGRI nor SPI can distinguish the emission of SN 2014J from the emission of any other source in M82. ISGRI, however, can easily differentiate between M82 and M81, which are separated by ~30′.

The energy resolution of ISGRI is ~10% at 100 keV. The ISGRI energy calibration uses the procedure implemented in OSA 10.0 (ref. 39). The images in broad energy bands were reconstructed using a standard mask–detector cross-correlation procedure, tuned to produce zero signal on the sky if the count rate across the detector matches the pattern expected from pure background, which was derived from the same data set by stacking detector images. The noise in the resulting images is fully consistent with the expected level, determined by photon counting statistics. The fluxes in broad bands were calibrated using Crab Nebula observations with INTEGRAL shortly before the observations of SN 2014J discussed here.

**Fitting 847 and 1,238 keV line parameters**

The flux, energy centroid and broadening of the lines were evaluated by fitting a Gaussian to portions of the SPI spectrum in the 800–900 and 1,100–1,350 keV bands. The best-fit parameters are given in Extended Data Table 1 along with 1 s.d. uncertainties. Although for plotting purposes we used heavily binned spectra (Figs 1 and 4), the fitting is done for the unbinned SPI spectrum in 0.5 keV-wide channels.

| Parameter | 847 keV line | 1238 keV line |
|---|---|---|
| Flux, $10^{-4}$ phot cm$^{-2}$ s$^{-1}$ | 2.34±0.74 | 2.78±0.74 |
| Luminosity, $10^{41}$ erg s$^{-1}$ | 4.7 | 8.1 |
| $V_{shift}$ (l.o.s. velocity), km s$^{-1}$ | -1900±1600 | -4300±1600 |
| Line width (l.o.s. velocity rms), $\sigma_v$, km s$^{-1}$ | 3600±1300 | 4700±1400 |

*Extended Data Table 1 Parameters of the observed gamma-ray lines of $^{56}$Co.*
*Each line is independently approximated with a Gaussian. Errors, 1 s.d.*

Because the decay time of $^{56}$Co ($\tau = 111.4$ days) and branching ratios (1 and 0.66 for the 847 and 1,238 keV lines, respectively) are known[25], it is straightforward to convert line fluxes into the mass of $^{56}$Co visible to INTEGRAL at the time of observation: $M_{Co} = (0.16 \pm 0.05)M_\odot$ and $(0.27 \pm 0.07)M_\odot$ for the 847 and 1,238 keV lines, respectively. These values can be considered model-independent lower limits on the amount of $^{56}$Co at day 75 since the explosion. The fractions of line photons escaping the ejecta without interactions were estimated from our fiducial model as 0.60 (847 keV) and 0.64 (1,238 keV). These values were used to correct the observed fluxes to derive estimates of the total $^{56}$Co mass at day 75, $M_{Co} = (0.26 \pm 0.08)M_\odot$ and $(0.42 \pm 0.11)M_\odot$, on the basis of the 847 and 1,238 keV line fluxes, respectively. The derived masses are consistent within the uncertainties with a mean value of $M_{Co} = (0.34 \pm 0.07)M_\odot$. Finally, a correction factor of 1/0.55 has been applied to convert the mass of $^{56}$Co at day 75 to the initial mass of $^{56}$Ni: $M_{Ni} = (0.6 \pm 0.13)M_\odot$. This factor accounts for time evolution of the Co mass in the decay chain from Ni to Fe.

In fully transparent ejecta, the centroid of emerging γ-ray lines should be unshifted (at least to first order in $V_e/c$, where $c$ is the speed of light). Opacity suppresses γ-rays coming from the receding part of the ejecta, leading to a blueshift of the visible line. Blueshift is indeed observed for both lines (Fig. 4). The corresponding mean velocity is $V_{Shift} = -3,100 \pm 1,100$ km s$^{-1}$. This value is slightly higher than the expected shift of $-1,280$ km s$^{-1}$, estimated from the fiducial model for γ-ray photons escaping the ejecta without interactions.

The expected line broadening (root mean squared line-of-sight velocity), $\sigma_r$, for transparent ejecta is directly related to the characteristic expansion velocity. In the model with an exponential density profile, $\sigma_r = 2V_e$. Indeed, in the model we found for directly escaping photons, $\sigma_r = 5,860$ km s$^{-1} \approx 2.1V_e$. The Gaussian fit to the observed lines yields a slightly lower value, $\sigma_r = 4,100 \pm 960$ km s$^{-1}$. It is possible that the exact values of the shift and broadening are affected by remaining uncertainties in the background model. However, the presence of shift and broadening at the level of few $10^3$ km s$^{-1}$ is very robust.

**Simplified model**

The emergent γ-ray spectrum from a type Ia supernova is determined by interactions of γ-ray quanta with the expanding ejecta and can be used as a proxy for the most basic properties of the supernova[15]. Although detailed modelling of the properties of SN 2014J is beyond the scope of this Letter, we use a simple model to qualitatively compare our results with expectations. Our basic observables are the line fluxes, primarily determined by $M_{Ni}$; line broadening, set by the typical energy release per unit mass; and the flux below 511 keV, which is sensitive to the Thompson depth of the ejecta. Accordingly, we build a spherically symmetric model of homologously expanding ejecta with mass $M_{Ej}$, ignoring the possible anisotropy of the $^{56}$Ni distribution expected in the scenario of white dwarf mergers[10]. The density follows an exponential law[26,27] $\rho \propto e^{-v/V_e}$, where $V_e = \sqrt{E_K/6M}$ and is truncated at $v = 10V_e$. Here $E_K$ is the kinetic energy of the ejecta. In this model a mass-weighted root-mean-squared velocity of the ejecta is $\sqrt{12}V_e$. All elements, including radioactive Ni and Co, are uniformly mixed through the entire ejecta, which are composed of iron group elements (58%) and Si and S in equal proportions (21% each). This is of course a strong simplification, certainly violated in the outer regions and in the core, but it allows us to specify the model completely with only three parameters: $M_{Ni}$, $M_{Ej}$ and $V_e$. The presence of the radioactive elements in the outer layers is crucial for the early phase of the supernova evolution, but becomes less important for days 50–100, which are the days of interest here. As a fiducial example, we use $M_{Ni} = 0.7 M_\odot$, $M_{Ej} = 1.38 M_\odot$ and $V_e = 2{,}800$ km s$^{-1}$, corresponding to $E_K = 1.3 \times 10^{51}$ erg.

A Monte Carlo radiative transfer code is used to calculate the emergent spectrum, which includes full treatment of Compton scattering (coherent and incoherent) and photoabsorption. Pair production by γ-ray photons is neglected. The positrons produced by $\beta^+$ decay of $^{56}$Co (19% of all decays) annihilate in place via positronium formation. Both two-photon annihilation into the 511 keV line and the ortho-positronium continuum are included. Our reference model was calculated for day 75 since the explosion. A time delay due to the finite propagation time of the photons is neglected (it amounts to few days from the radius where the bulk of the mass is located).

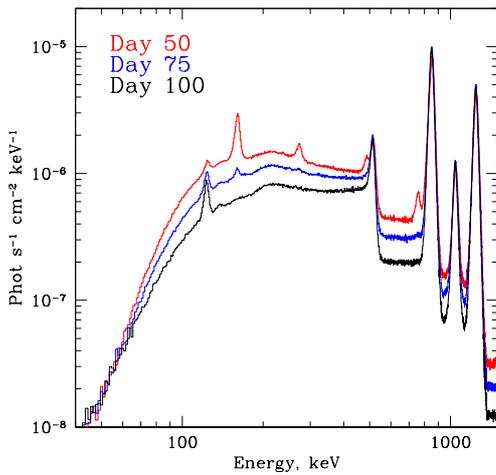

*Extended Data Figure 3 Predicted spectra for days 50, 75 and 100 after explosion. The 3PAR model spectrum calculated for day 75 is used for comparison with the INTEGRAL data obtained between day 50 and 100 since the explosion. Weak lines below 200 keV correspond to $^{56}$Ni (day 50) and $^{57}$Co (day 100).*

Examples of the model spectra for days 50, 75 and 100 are shown in Extended Data Fig. 3. To a first approximation (justified by the low signal-to-noise ratio of the observed spectrum), the model spectrum for day 75 can be used for comparison with the observed spectrum, accumulated by INTEGRAL over days 50–100. The contributions of various components to the total spectrum are shown in Extended Data Fig. 4. The most prominent are the lines at 847 and 1,238 keV. These lines are broadened by the expansion velocity of the ejecta and escape without any interactions (the role of coherent scattering is negligible at these energies). The line shape is also modified by opacity effects, because γ-ray photons produced in the nearest (approaching) side of the ejecta have a higher chance of reaching an observer. This effect causes the line shapes to be skewed towards the blue side. The scattered continuum associated with most prominent lines extends down to energies ~100 keV. At lower energies, photoabsorption becomes dominant and the flux drops. In addition, three-photon annihilation, associated with the decay of ortho-positronium, produces a significant contribution to the continuum flux. Unlike the scattered continuum, this component does not scale with the Thomson depth of the ejecta and becomes progressively more important for late evolution of the emergent spectrum.

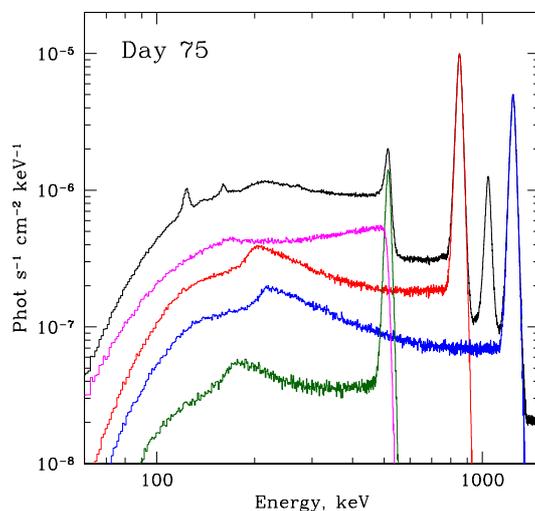

***Extended Data Figure 4 Contributions of various components to the model spectrum.*** *The lines are formed by γ-rays escaping the ejecta without interactions. The low-energy tail of each line is due to Compton down-scattering of the photons because of the recoil effect. The 'humps' in the tails correspond to the scattering by 180°. The magenta line shows the contribution of the ortho-positronium annihilation. Annihilation of para-positronium contributes to 511 keV line.*

With our simple three-parameter model it is possible to run a grid of models to evaluate a plausible range of basic parameters. This was done by varying $M_{Ni}$, $M_{Ej}$ and $V_e$, comparing the model with the spectrum and calculating $\chi^2$. It is convenient to express the success of the model by the reduction of $\Delta\chi^2$ relative to the null hypothesis of no source. The null hypothesis gives $\chi^2 = 1{,}948.7$ for 1,905 degrees of freedom for the combined SPI+ISGRI spectrum. The best-fitting three-parameter model has $\Delta\chi^2 = 50.5$, corresponding to a 7.1 s.d. detection with $M_{Ni} = 0.56\,M_\odot$, $M_{Ej} = 1.2\,M_\odot$ and $V_e = 3{,}000$ km s$^{-1}$, corresponding to $E_K = 1.3 \times 10^{51}$ erg. The same model allows for calculation of 1 s.d. confidence intervals (for a single parameter of interest) by identifying parameter space, which has $\Delta\chi^2$ smaller by 1 than the best-fitting model. Corresponding confidence intervals are shown in Extended Data Fig. 5.

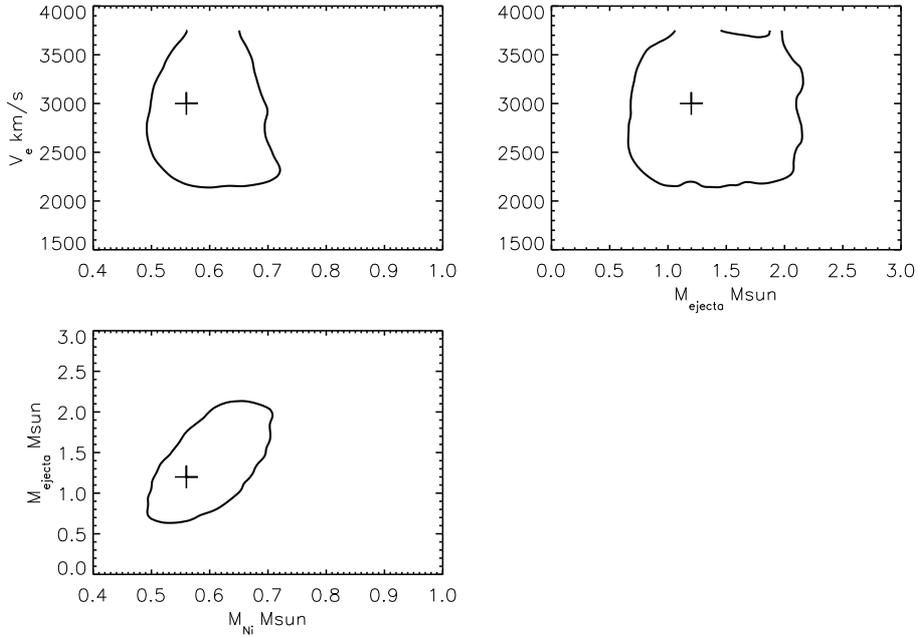

***Extended Data Figure 5 Confidence contours for our three-parameter model.*** *The cross shows the best-fit values. Contours are plotted at $\Delta\chi^2 = 1$ with respect to the best-fit value and characterize the 1 s.d. confidence interval for a single parameter of interest.*

**Specific explosion models**

Apart from our simplified three-parameter model, we compare the spectrum (SPI+ISGRI) with the expected spectra calculated (E.B. *et al.*, manuscript in preparation) for several detailed type Ia supernova explosion models. For each model, we calculate the reduction of $\Delta\chi^2$ relative to the null hypothesis of no source. The resulting values and basic characteristics of the models are given in Extended Data

Table 2. The set of models includes the classic W7 (ref. 2) and DD4 (ref. 30) models, the pure-detonation model, DETO[29], the sub-Chandrasekhar model, HED6 (ref. 30), our three-parameter model 3PAR with fiducial and best-fitting parameters, and several variants of the delayed-detonation model DDT1p (E.B. *et al.*, manuscript in preparation). DDT1p4 was built to match SN 2014J in the visible band. DDT1p1 is a slightly less energetic version of DDT1p4. In the DDT1p4halo model, the object is surrounded by a $0.2M_\odot$ 'halo', which can be envisaged in a slow merger scenario.

| Model | $M_{Ni}$, $M_{Sun}$ | $M_{Ej}$, $M_{Sun}$ | $E_K$, $10^{51}$ erg | $\Delta\chi^2$ |
|---|---|---|---|---|
| W7[1] | 0.59 | 1.38 | 1.24 | 54.4 |
| DDT1p1 | 0.54 | 1.36 | 1.29 | 52.5 |
| DD4[30] | 0.61 | 1.39 | 1.24 | 52.0 |
| DDT1p4 | 0.66 | 1.36 | 1.35 | 51.9 |
| 3PAR, best-fitting | 0.56 | 1.20 | 1.3 | 50.5 |
| 3PAR, fiducial | 0.70 | 1.38 | 1.3 | 49.3 |
| DDT1p4halo | 0.62 | 1.55 | 1.3 | 49.1 |
| HED6[29] | 0.26 | 0.77 | 0.72 | 38.2 |
| DETO[28] | 1.16 | 1.38 | 1.44 | 12.1 |

*Extended Data Table 2 Comparison of typical type Ia supernova explosion models with the data. The rightmost column shows the improvement in $\chi^2$, relative to the null hypothesis of no source, by assuming a type Ia supernova spectral model (no free parameters). The null hypothesis gives $\chi^2 = 1,948.7$ for 1,905 degrees of freedom.*

Although the single WD models discussed above provide consistent description of the INTEGRAL data, these observations by themselves do not immediately exclude a double WD merger scenarios, provided that similar amount of $^{56}$Ni is synthesized. In this regard, we note that no direct evidence that the progenitor of SN2014J was a single accreting WD has been found so far, placing tight limits on the most popular accretion scenarios[40,22,41,42].

**Optical bolometric luminosity during INTEGRAL observations.** Using recent BVRIJHK photometry[43], we estimated the supernova optical bolometric luminosity on day 75 after explosion (median of INTEGRAL observations). For $A_V$=1.85 mag and $R_V$=1.6 we found the bolometric luminosity of ~1.1 $10^{42}$ erg s$^{-1}$, which agrees well with the estimates of deposited power in our fiducial model.